\documentclass[12pt,aps,preprint]{revtex4}
\usepackage{amssymb,graphics}

\begin{document}

\title{Radiation pressure on a dielectric boundary: was Poynting wrong?}
\author{A.~Hirose and R.~Dick}
\affiliation{Department of Physics and Engineering Physics, University of Saskatchewan \\
116 Science Place, Saskatoon, SK S7N 5E2, Canada}

\begin{abstract}
{When a plane electromagnetic wave in air falls on a flat dielectric
boundary, the dielectric body is pulled toward the air as predicted by
Poynting a century ago. According to Noether's theorem, the momentum in
the direction parallel to the boundary must be conserved among the incident,
reflected and transmitted waves. This uniquely determines the expression for
the wave momentum in the dielectric medium which agrees with the Minkowski
form. The inward force recently predicted and accompanied by Abraham's
momentum are not consistent with Fresnel's formulae and basic symmetry
principles.}
\end{abstract}

\maketitle

\narrowtext

The problem of momentum conservation in reflection and transmission of an
electromagnetic wave entering (or leaving) a dielectric medium has been a
matter of controversy over the past century. In 1905,
Poynting \cite{Poynting} predicted that the dielectric is pulled toward
air region. In
this case, the momentum of the wave transmitted into the dielectric
coincides with that proposed by Minkowski \cite{Minkowski} which is also
consistent with the classical relationship between the energy flux and
momentum flux,
\begin{equation}
\text{energy flux = wave velocity }\times \text{ momentum flux}
\label{classical}
\end{equation}%
(We consider nondispersive waves since dispersion does not play important
roles in the argument to follow.) Poynting even performed experiments using
internal reflections of visible light in a glass slab and measured a torque
exerted by the light wave in favorable agreement with the prediction. The
outward force is also consistent with what is expected from the Maxwell's
stress tensor. In the case of normal incidence, the electric field is
tangential to the surface and thus should be continuous across the boundary.
Then there is a pressure differential 
\[
\frac{1}{2}\left( \varepsilon -\varepsilon _{0}\right) E_{t}^{2} 
\]%
acting from the dielectric region to air region (from higher energy density
region to the lower). Momentum conservation among the three waves (incident,
reflected , and transmitted) does not hold and the extra mechanical
momentum, which is absorbed by the dielectric medium, exactly coincides with
the pressure difference above \cite{Kemp}. Momentum non-conservation in wave
reflection and transmission is not surprising and has been noticed in
mechanical waves as well \cite{Rowland}.

The validity of Poynting's analysis has recently been questioned by Loudon 
\cite{Loudon} who re-analyzed the momentum transfer using the Lorentz force
arising from the polarization current. Without assuming a priori knowledge
of the wave momentum in a dielectric, he concluded that the momentum
absorbed by the dielectric medium is equal in magnitude but exactly opposite
to the force predicted by Poynting. In Loudon's theory, the momentum carried
by the wave in the dielectric turned out to be that proposed by Abraham \cite%
{Abraham} and smaller than that of Minkowski by a factor $n^{2}=\varepsilon
/\varepsilon _{0}$ (in the case of nonmagnetic dielectric). Since the energy
flux density is the same in both formulations because of energy
conservation, the consequence is that the classical relationship in Eq. (1)
has to be discarded in the Abraham's theory. Despite these difficulties in
the Abraham's theory, the issue still remains unsettled because no basic
laws have been identified to uniquely define the momentum of electromagnetic
waves in a dielectric medium. In some theory \cite{Mansripur}, an algebraic
average of the two momenta has been claimed to be the wave momentum in
dielectric materials. Moller \cite{Moller} stated that \textquotedblleft one
is free to make a convenient choice since in any case the total moemtum
density of fields and matter satsifies a conservation
law.\textquotedblright\ We believe there must be a basic law that uniquely
determines the expression of the wave momentum in a dielectric which will
prove Poynting was either correct or wrong.

In this Letter, we revisit the problem from an entirely different
perspective exploiting the Noether theorem \cite{Noether} which requires
that momentum be conserved in the spatial direction of translational
invariance. In the case of oblique incidence of electromagnetic wave on a
dielectric medium, the Noether theorem tells us that the Poynting energy
flux densities in the normal direction is conserved (consequence of
invariance in the direction of time axis) and the momenta of the incident,
reflected and transmitted waves parallel to the boundary surface are
conserved. As will be shown, the latter constraint \emph{uniquely}
determines the momentum of electromagnetic waves in a dielectric medium
which agrees with that of Minkowski. The parallel momentum conservation has
not been considered or utilized (to our knowledge) in the past in terms of
the fundamental symmetry properties of physical systems (Noether's theorem)
which may have contributed to the ambiguity in defining wave momentum in a
dielectric medium. The Fresnel's formulae \cite{Stratton} for reflected and
transmitted waves are fully consistent with the Noether theorem and unique
definition of the Minkowski momentum is actually hidden in the set of
Fresnel's formulae.

Let us first review briefly the underlying problem by considering a very
simple case: a plane electromagnetic wave having an amplitude $E_{0}$ in air
falling normal on a flat surface of a semi-infinite, non-magnetic dielectric
medium with an index of refraction $n.$ (This is actually the geometry used
in most studies.) The amplitudes of reflected and transmitted waves can be
readily found from the continuity of electric and magnetic fields at the
boundary and are given by 
\begin{equation}
E_{r}=rE_{0}=\frac{1-n}{1+n}E_{0},\text{ }E_{t}=\left( 1+r\right) E_{0}=%
\frac{2}{1+n}E_{0}
\end{equation}%
The energy flux densities (Poynting vectors) of the incident, reflected and
transmitted waves are well known, 
\begin{equation}
S_{i}=c\varepsilon _{0}E_{0}^{2},\text{ }S_{r}=\left( \frac{1-n}{1+n}\right)
^{2}c\varepsilon _{0}E_{0}^{2},\text{ }S_{t}=\left( \frac{2n}{1+n}\right)
^{2}\frac{c}{n}\varepsilon _{0}E_{0}^{2}
\end{equation}%
It is evident that energy conservation holds $S_{i}=S_{r}+S_{t}.$ However,
momentum flux density is not conserved among the three waves. According to
the classical relationship in Eq. (\ref{classical}), the momentum flux
densities of the three waves are%
\begin{equation}
P_{i}=\varepsilon _{0}E_{0}^{2},\text{ }P_{r}=-\left( \frac{1-n}{1+n}\right)
^{2}\varepsilon _{0}E_{0}^{2},\text{ }P_{t}=\left( \frac{2}{1+n}\right)
^{2}\varepsilon E_{0}^{2}
\end{equation}%
The momentum conservation clearly does not hold among the three waves and
the difference 
\begin{equation}
\Delta P=P_{i}-P_{r}-P_{t}=\frac{2\left( n^{2}+1\right) }{\left( 1+n\right)
^{2}}\varepsilon _{0}E_{0}^{2}-\left( \frac{2n}{1+n}\right) ^{2}\varepsilon
_{0}E_{0}^{2}=-\frac{2\left( n^{2}-1\right) }{(n+1)^{2}}\varepsilon
_{0}E_{0}^{2}
\end{equation}%
must be acting on the dielectric body as a mechanical force. Since the
electric field at the boundary $E_{t}=2E_{0}/(1+n)$ is continuous, the
difference in the energy densities exerts a pressure which exactly coincides
with the extra momentum flux density, 
\begin{equation}
\frac{1}{2}\left( \varepsilon _{0}-\varepsilon \right) E_{t}^{2}=\Delta P
\label{stress}
\end{equation}%
It is noted that the momentum flux density adopted in the analysis here is
that of Poynting-Minkowski. Loudon \cite{Loudon} showed an alternative
partition, 
\begin{equation}
P_{i}-P_{r}=\frac{2\left( n^{2}+1\right) }{\left( 1+n\right) ^{2}}%
\varepsilon _{0}E_{0}^{2}=\frac{4}{\left( 1+n\right) ^{2}}\varepsilon
_{0}E_{0}^{2}+\frac{2\left( n^{2}-1\right) }{(n+1)^{2}}\varepsilon
_{0}E_{0}^{2}
\end{equation}%
Here the partition is between the Abraham's momentum%
\[
P_{A}=\frac{4}{\left( 1+n\right) ^{2}}\varepsilon _{0}E_{0}^{2} 
\]%
and the \textquotedblleft surface force\textquotedblright 
\[
\frac{2\left( n^{2}-1\right) }{(n+1)^{2}}\varepsilon _{0}E_{0}^{2}=\frac{1}{2%
}\left( \varepsilon -\varepsilon _{0}\right) E_{t}^{2} 
\]%
which is inward. However, as will be shown, Abraham's momentum cannot
satisfy the Noether theorem.

The Noether theorem pertains to conservation of energy and linear momentum
in the direction parallel to the boundary. We therefore consider oblique
incidence with incidence angle $\alpha $ and refraction angle $\alpha
^{\prime }$. The angles $\alpha $ and $\alpha ^{\prime }$ are related
through the Snell's law, $\sin \alpha =n\sin \alpha ^{\prime }.$ If the
magnetic field is in the incidence plane, the electric fields are parallel
to the surface. For an incident electric field $E_{0},$ reflected and
transmitted fields are%
\[
E_{r}=rE_{0},\text{ }E_{t}=\left( 1+r\right) E_{0} 
\]%
where $r$ is the reflection coefficient for the assumed polarization%
\begin{equation}
r=\frac{\sin \left( \alpha ^{\prime }-\alpha \right) }{\sin \left( \alpha
^{\prime }+\alpha \right) }  \label{reflection}
\end{equation}%
The Poynting fluxes associated with each wave are%
\[
\text{ incident: }S_{i}=c\varepsilon _{0}E_{0}^{2};\text{ reflected: }%
S_{r}=r^{2}c\varepsilon _{0}E_{0}^{2} 
\]%
\[
\text{transmitted: }S_{t}=(1+r)^{2}\frac{c}{n}n^{2}\varepsilon
_{0}E_{0}^{2}=(1+r)^{2}cn\varepsilon _{0}E_{0}^{2} 
\]%
The normal components of the Poynting fluxes are conserved%
\begin{equation}
(S_{i}-S_{r})\cos \alpha =S_{t}\cos \alpha ^{\prime }
\end{equation}%
since for the reflection coefficient in Eq. (\ref{reflection}), the
following identity holds%
\begin{equation}
\left( 1-r\right) \cos \alpha =n\left( 1+r\right) \cos \alpha ^{\prime }
\label{E conservation}
\end{equation}%
This can be interpreted as energy flux (not density) along the direction of
respective waves since $\cos \alpha $ factor essentially indicates the beam
width. (See Fig. 1.) Then according to Eq. (1), the classical momentum
fluxes (not flux density) of each wave are%
\begin{equation}
\text{incident: }P_{i}=\varepsilon _{0}E_{0}^{2}\cos \alpha ;\text{
reflected: }P_{r}=r^{2}\varepsilon _{0}E_{0}^{2}\cos \alpha
\end{equation}%
\begin{equation}
\text{transmitted: }P_{t}=(1+r)^{2}n^{2}\varepsilon _{0}E_{0}^{2}\cos \alpha
^{\prime }
\end{equation}%
The components parallel to the boundary surface are 
\begin{equation}
\text{incident: }\varepsilon _{0}E_{0}^{2}\cos \alpha \sin \alpha ;\text{
reflected: }r^{2}\varepsilon _{0}E_{0}^{2}\cos \alpha \sin \alpha
\end{equation}%
\begin{equation}
\text{transmitted: }(1+r)^{2}n^{2}\varepsilon _{0}E_{0}^{2}\cos \alpha
^{\prime }\sin \alpha ^{\prime }
\end{equation}%
It is evident that momentum conservation holds,%
\begin{equation}
\varepsilon _{0}E_{0}^{2}\left( 1-r^{2}\right) \cos \alpha \sin \alpha
=(1+r)^{2}n^{2}\varepsilon _{0}E_{0}^{2}\cos \alpha ^{\prime }\sin \alpha
^{\prime }  \label{para momentum}
\end{equation}%
if Snell's law $\sin \alpha ^{\prime }=\sin \alpha /n$ and energy
conservation in Eq. (\cite{E conservation}) are recalled. Therefore,
Fresnel's reflection/transmission formulae fully satisfy the Noether
theorem. Parallel momentum conservation is a consequence of energy
conservation as well. From the last term in Eq. (\ref{para momentum}), it is
concluded that the momentum flux density of electromagnetic wave in a
dielectric medium is \emph{uniquely} given by%
\begin{equation}
n^{2}\varepsilon _{0}E_{t}^{2}=\varepsilon E_{t}^{2}
\end{equation}%
Other forms of momentum such as Abraham's momentum do not satisfy the basic
theorem. The momentum agrees with Minkowski's definition of photon momentum
in a dielectric and is also implicitly consistent with the negative pressure
predicted by Poynting.

When the electric field is in the incidence plane, the reflected and
transmitted waves are%
\[
E_{r}=\frac{\tan \left( \alpha -\alpha ^{\prime }\right) }{\tan \left(
\alpha +\alpha ^{\prime }\right) }E_{0},\text{ }E_{t}=\frac{2\cos \alpha
\sin \alpha ^{\prime }}{\sin \left( \alpha +\alpha ^{\prime }\right) \cos
\left( \alpha -\alpha ^{\prime }\right) }E_{0} 
\]%
In this case too, energy conservation 
\[
E_{0}^{2}\left( 1-r^{2}\right) \cos \alpha =nE_{t}^{2}\cos \alpha ^{\prime } 
\]%
holds and we find $n^{2}\varepsilon _{0}E_{t}^{2}$ to be the wave momentum
flux in the dielectric.

The Minkowski momentum is therefore arises from the translational invariance
and ensuing conservation of momentum parallel to the boundary surface. The
Noether theorem asserts that translational invariance of a physical system
in a spatial direction implies momentum conservation in this direction. The
Noether theorem also tells us that invariance under time translations
implies energy conservation. For a stationary planar air-dielectric
interface, we therefore must have momentum conservation parallel to the
interface and energy conservation. The analyses for separate polarizations
presented above can be unified as follows.

We consider incidence of an electromagnetic wave with incidence angle $%
\alpha $ with respect to the normal axis as shown in Fig. 1. The region $z<0$
is assumed to be air with $n=1$ and the region $z>0$ is a medium with
permittivity $\varepsilon $ and permeability $\mu .$ The reflected and
transmitted waves in terms of the incident fields $E_{i\parallel }$ and $%
E_{i\perp }$ with polarizations parallel and perpendicular to the incidence
plane are%
\begin{equation}
\frac{E_{r\parallel }}{E_{i\parallel }}=\frac{\sqrt{\varepsilon _{0}\mu }%
\cos \alpha ^{\prime }-\sqrt{\varepsilon \mu _{0}}\cos \alpha }{\sqrt{%
\varepsilon _{0}\mu }\cos \alpha ^{\prime }+\sqrt{\varepsilon \mu _{0}}\cos
\alpha }  \label{A}
\end{equation}%
\begin{equation}
\frac{E_{t\parallel }}{E_{i\parallel }}=\frac{2\sqrt{\varepsilon _{0}\mu }%
\cos \alpha }{\sqrt{\varepsilon _{0}\mu }\cos \alpha ^{\prime }+\sqrt{%
\varepsilon \mu _{0}}\cos \alpha }  \label{B}
\end{equation}%
\begin{equation}
\frac{E_{r\perp }}{E_{i\perp }}=\frac{\sqrt{\varepsilon _{0}\mu }\cos \alpha
-\sqrt{\varepsilon \mu _{0}}\cos \alpha ^{\prime }}{\sqrt{\varepsilon
_{0}\mu }\cos \alpha +\sqrt{\varepsilon \mu _{0}}\cos \alpha ^{\prime }}
\label{C}
\end{equation}%
\begin{equation}
\frac{E_{t\perp }}{E_{i\perp }}=\frac{2\sqrt{\varepsilon _{0}\mu }\cos
\alpha }{\sqrt{\varepsilon _{0}\mu }\cos \alpha +\sqrt{\varepsilon \mu _{0}}%
\cos \alpha ^{\prime }}  \label{D}
\end{equation}%
Energy conservation and the well known expression for the wave energy
density $u=\varepsilon \mathbf{E}^{2}$ in an isotropic medium then imply%
\begin{equation}
\varepsilon _{0}\mathbf{E}_{i}^{2}V=\varepsilon _{0}\mathbf{E}%
_{r}^{2}V+\varepsilon \mathbf{E}_{t}^{2}V^{\prime }
\label{energy conservation}
\end{equation}%
where%
\[
\mathbf{E}_{i}=\mathbf{E}_{i\parallel }+\mathbf{E}_{i\perp },\mathbf{E}_{r}=%
\mathbf{E}_{r\parallel }+\mathbf{E}_{r\perp },\mathbf{E}_{t}=\mathbf{E}%
_{t\parallel }+\mathbf{E}_{t\perp } 
\]%
and 
\begin{equation}
\frac{V^{\prime }}{V}=\frac{\cos \alpha ^{\prime }}{n\cos \alpha }=\frac{1}{n%
}\sqrt{\frac{1-\frac{1}{n^{2}}\sin ^{2}\alpha }{1-\sin ^{2}\alpha }}
\label{volume}
\end{equation}%
is the volume conversion factor which results from the dilation of the beam
width $\cos \alpha ^{\prime }/\cos \alpha $ upon entry to the dielectric and
longitudinal compression of the incident beam due to the velocity ratio $%
c^{\prime }/c=1/n.$ The condition for energy conservation between the
incident, reflected and transmitted beams therefore reads%
\begin{equation}
\varepsilon _{0}\left( \mathbf{E}_{i}^{2}-\mathbf{E}_{r}^{2}\right) \cos
\alpha =\frac{1}{n}\varepsilon \mathbf{E}_{t}^{2}\cos \alpha ^{\prime }
\label{conservation2}
\end{equation}%
which is equivalent to the conservation of normal components of the Poynting
vectors as noted earlier. It is easily confirmed from Eqs. (\ref{A}) - (\ref%
{D}) that the condition is identically fulfilled for arbitrary incidence
angle $\alpha $ and polarization, as was of course expected. However, we
have to recall that this little calculation is to confirm the volume
conversion factor in Eq. (\ref{volume}), and to remind the reader that $%
u_{t}=\varepsilon \mathbf{E}_{t}^{2}$ is considered as the energy density in
the transmitted wave in the dielectric without (to our knowledge) anybody
ever questioning whether this would have to be split into a genuine
electromagnetic component and a co-moving material component, or whether it
would have to be denoted as a \textquotedblleft pseudo-energy
density\textquotedblright\ of the transmitted wave.

Another well known statement of energy conservation between the incident,
reflected and transmitted beams results from Eq. (\ref{conservation2}) if we
take into account the relation between energy density and energy flux
density (Poynting flux) in a medium with refractive index $n,$ 
\[
\left\vert \mathbf{S}\right\vert =\frac{c}{n}\varepsilon \mathbf{E}^{2} 
\]%
Multiplying Eq. (\ref{energy conservation}) by $c$ yields%
\[
\left( S_{i}-S_{r}\right) \cos \alpha =S_{t}\cos \alpha ^{\prime } 
\]%
Multiplying further by $\sin \alpha ,$ we obtain%
\[
\left( S_{ix}-S_{rx}\right) \cos \alpha =nS_{tx}\cos \alpha ^{\prime } 
\]%
where $S_{jx}$ is the $x$ component of each Poynting flux. Multiply by $%
V/\left( c^{2}\cos \alpha \right) =nV^{\prime }/\left( c^{2}\cos \alpha
^{\prime }\right) $ to obtain%
\begin{equation}
\frac{1}{c^{2}}\left( S_{ix}-S_{rx}\right) V=\frac{n^{2}}{c^{2}}%
S_{tx}V^{\prime }
\end{equation}%
This asserts that the momentum density of the transmitted wave in the
dielectric medium is uniquely determined as 
\begin{equation}
\mathbf{p}=\frac{n^{2}}{c^{2}}\mathbf{S}
\end{equation}%
which is the Minkowski's formula.

In conclusion, in wave reflection and transmission of electromagnetic wave
at a dielectric boundary, the wave momentum along the boundary surface must
be conserved according to Noether's theorem. The Fresnel's formula are
consistent with this basic requirement. The momentum flux density and
momentum of electromagnetic wave in a dielectric medium uniquely takes the
Minkowski's form without room for alternatives. The classical relationship
between the energy flux density and momentum flux density in Eq. (1) is
preserved to our comfort. The force to act on a dielectric when a wave
enters it is outward as predicted and observed by Poynting, and as also
required from Maxwell's stress tensor.

\noindent

Helpful communication with Professors R. Loudon and M. Mansripur is
acknowledged with gratitude.

This research is sponsored by the Natural Sciences and Engineering Research
Council of Canada and by Canada Research Chair Program.

\newpage

\begin{figure}
\includegraphics{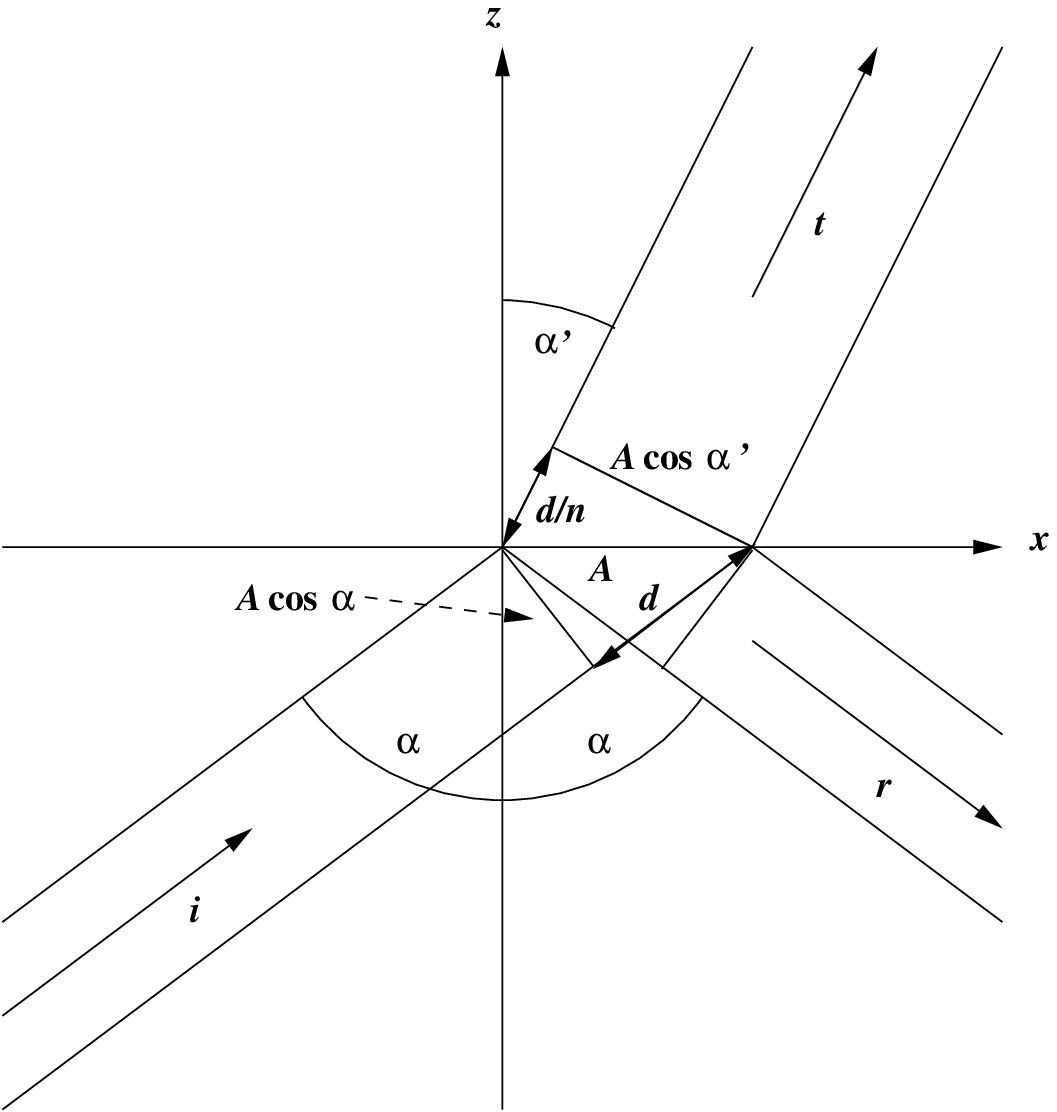} 
\caption{The lower half space $z<0$ is vacuum or air with dielectric
parameters $\protect\epsilon _{0}$ and $\protect\mu _{0}$. The upper half
space $z>0$ is dielectric material with parameters $\protect\epsilon $ and $%
\protect\mu $. The letters i, r, and t denote the incoming, reflected and
transmitted beam. The beam cross sections $A\cos \protect\alpha $ and $A\cos 
\protect\alpha ^{\prime }$, and the lengths $d$ and $d^{\prime }=d/n$ are
displayed to explain the volume conversion factor $V^{\prime }/V$ between
the incoming and transmitted waves.}
\label{Fig 1}
\end{figure}


\begin{thebibliography}{99}
\bibitem{Kemp} B. A. Kemp, T. M. Grzegorczyk, and J. A. Kong, Opt. Express 
\textbf{13}, 9280 (2005). In this paper, it was correctly pointed out that
the outward force due to the stress tensor corresponds only to the
mechanical momentum absorbed by the dielectric body. It should be contained
as one part in the momentum based on the Lorentz force which has not been
calculated properly in this paper.
\end{thebibliography}
\end{document}